\documentclass[amssymb,aps,prli,preprint]{revtex4}
\usepackage{latexsym,amsmath,graphics,graphicx,cmbright,exscale,colordvi}
\def\be{\begin{equation}}
\def\ee{\end{equation}}
\def\beq{\begin{eqnarray}}
\def\eeq{\end{eqnarray}}

\begin{document}
\renewcommand{\familydefault}{\sfdefault}
\renewcommand{\sfdefault}{cmbr}

\title{Theory of Local Dynamical Magnetic Susceptibilities from the Korringa-Kohn-Rostoker Green Function Method}
\author{S. Lounis$^1$}\email{slounis@uci.edu}
\author{A. T. Costa$^{1,2}$}
\author{R. B. Muniz$^2$}
\author{D. L. Mills$^1$}
\affiliation{$^1$ Department of Physics and Astronomy, University of California Irvine, California, 92697 USA}
\affiliation{$^2$ Instituto de F\'isica, Universidade Fedeal Fluminense, 
24210-340 Niter\'oi, RJ, Brazil}

\begin{abstract}
Within the framework of time-dependent density functional theory combined with the Korringa-Kohn-Rostoker Green function formalism, we present a real space methodology to investigate dynamical magnetic excitations from first-principles. We set forth a scheme which enables one to deduce the correct effective Coulomb potential needed to preserve the spin-invariance signature in the dynamical susceptibilities, i.e. the Goldstone mode. We use our approach to explore the spin dynamics of 3d adatoms and different dimers deposited on a Cu(001) with emphasis on their decay to particle-hole pairs.
\end{abstract}
\maketitle
\date{\today}
%\pacs{}

%\begin{multicols}{2}
%narrowtext
\section{Introduction}

The magnetic functionalization of nanostructures made of few atoms 
requires the understanding of spin-excitations 
at the nanoscale and subnanoscale level. Recently, state of the art experiments based on 
scanning tunneling microscopy (STM) were utilized to 
excite and control the magnetic states of single adatoms sitting on 
semi-insulating~\cite{heinrich} or metallic~\cite{balashov,khajetoorians} surfaces. 
The spin dynamics of moment bearing 3d metal atoms have been 
probed in those experiments but often the theoretical picture used for the 
interpretation is based on a model Hamiltonian describing an atomic 
like localized moment with integer or half-integer spin. Such a model is useful only for systems where the substrate interacts weakly with the 
adsorbate~\cite{heinrich}; it fails qualitatively to describe cases with 
strong coupling to the substrate electrons where hybridization leads to moments far from integer and half integer values, and d levels with widths that can range from a few hundred millivolts to perhaps an electron volt. This paper presents a scheme wherein one may address the commonly encountered strongly coupled systems, with density functional theory as the basis.  In contrast to empirical tight binding schemes used earlier~\cite{cooke,mills,muniz,costa} the method set forth in this paper incorporates a proper ab-initio based description of the one electron physics from upon which our description of spin dynamics is erected. Also the scheme set forth in this paper may be implemented with modest computational labor.  

Several approaches have been proposed to describe inelastic STM experiments involving 
the above mentioned local moment picture~\cite{fransson,balatsky,fernandez,lorente,persson} 
but none are based on taking full account 
of the electronic structure of the adsorbates as well as the substrates 
including the effects of hybridization. The latter requires, among 
other ingredients~\cite{future}, the 
evaluation of the transverse magnetic response function $\chi$ or the so-called transverse 
dynamical magnetic susceptibility that relates, in linear response theory, the amplitude of  
 the transverse spin motion $m_{x,y}$ produced by a transverse 
external magnetic field $B_{ext}$ of frequency $\omega$. There are three major roads followed to compute 
$\chi$: (i) empirical tight-binding theory (ETB)~\cite{cooke,mills,muniz,costa}, (ii) 
time-dependent density functional theory (TD-DFT)~\cite{gross1,stenzel,savrasov,staunton,tddft_book,buczek,niesert,lounis}, and (iii) many body perturbation theory (MBPT) using 
the Random Phase Approximation (RPA) and DFT~\cite{aryasetiawan,sasioglu}. The calculation of $\chi$ requires one 
to solve a Dyson equation whose solution may be written schematically in the form:
\beq
\chi = \chi_0 (1- U \chi_0)^{-1}
\label{dyson}
\eeq
As noted in Ref.~\cite{lounis}, $\chi_0$ is described by a different 
nomenclature depending on the method used to calculate it. Within TD-DFT~\cite{gross1,tddft_book}, 
$\chi_0$ is known as the Kohn-Sham susceptibility and $U$ is the exchange and correlation 
kernel that if ideally known completely would render Eq.~\ref{dyson} the exact solution. $U$ is obviously 
approximated in practice, for example, by the adiabatic local spin density approximation. 
It turns out that evaluating Eq.~\ref{dyson} is computationally very challenging, especially within the TD-DFT or the MBPT. This explains  
the very few calculations found in the literature, almost all of which address bulk systems. This makes it even more challenging to simulate inelastic STM experiments that examine adatoms deposited on surfaces.
  Recently, we developed a method~\cite{lounis} that handles the calculation of the 
transverse dynamical magnetic susceptibility in a scheme that resembles ETB but is based on TD-DFT. Thus the method incorporates   
full self-consistent first-principles calculations of the underlying electronic structure. Two interesting results were obtained: 
(i) a justification of the Lowde and Windsor scheme\cite{lowde} 
emerged from the analysis and (ii) values of $U$ determined from first-principles 
for different systems are in accordance with the empirical values extracted 
from photoemission data by Himpsel.~\cite{himpsel}

In our previous paper~\cite{lounis}, we addressed a central question related to the 
practical determination of $\chi_0$ and $U$, within the framework of density functional based schemes. It is known, but often not discussed explicitly, that the Goldstone theorem is not satisfied, in practice, when 
solving Eq.~\ref{dyson} within TD DFT schemes. We remark that within the framework of the empirical tight binding scheme, the Goldstone theorem is satisfied exactly, as demonstrated earlier~\cite{muniz}. The Goldstone 
theorem, when satisfied, insures that the zero wave
 vector spin waves have precisely zero frequency (when spin orbit 
coupling is set aside).
The reason the Goldstone theorem is not satisfied within density functional 
based schemes is that the numerical methods used to 
extract $U$ and $\chi_0$ are not compatible with the Ward identity. To compensate for this problem, 
Sasioglu {\it et al.}~\cite{sasioglu} correct $U$ by 45\% in their study of bulk Ni while Buczek {\it et al.} find a finite frequency for the Goldstone mode~\cite{buczek}. 
To cure such inconsistencies, an ad-hoc shift by hand 
of the value of $U$ is used commonly. Our aim is to demonstrate that such corrections could be dangerous, for instance, 
when the system under investigation contains more than two atoms in the unit cell.
In Ref.~\cite{lounis}, we set forth and utilized a sum rule that allows one 
to generate a $U$ that is fully compatible with the Goldstone mode.

The discussion of the sum rule in ref.~\cite{lounis} was brief, though its application was illustrated. In this paper, we provide a detailed derivation of our scheme~\cite{lounis} including the sum rule needed to determine $U$. Our method is based on the Korringa-Kohn-Rostoker single particle Green function (KKR-GF)~\cite{KKR} which contains an ab-initio description of the electronic structure.

We remark that in earlier work, the empirical tight binding method has been used successfully to describe spin waves in films on substrates~\cite{mills,costa} along with the spin dynamics of adatoms as probed in the recent STM experiments~\cite{muniz}. In this approach, it is necessary to make contact with electronic structure calculations for the purpose of extracting the tight binding parameters required to describe the one electron properties of the system of interest. Often appropriate electronic structure calculations are unavailable, or if they are it can be a challenge to extract appropriate parameters in an unambiguous manner for complex systems such as ultrathin films adsorbed on substrates. The approach we develop here eliminates this issue completely, while at the same time it provides a computationally straightforward scheme for generating the dynamic transverse susceptibility. 	

\section{Structure of the Theory; The Sum Rule and the Effective $U$\label{section_sumrule}}

It is, of course, possible in principle to calculate the Kohn-Sham non interacting susceptibility $\chi_0$. In this section, we show that once 
$\chi_0$ assumed known, we can derive a prescription for generating the effective Coulomb interaction $U$ which enters Eq.~\ref{dyson} that is 
fully compatible with the Goldstone theorem. In effect, $U$ is a functional 
of $\chi_0$. With $U$ determined in the manner we describe, there is no reason 
for ad-hoc adjustment of this central parameter. We also describe a scheme which allows one to generate a physically sensible approximation to $\chi_0$ 
that is straightforward and simple to implement. We then use this scheme to generate a series of explicit predictions regarding the nature of spin excitations of adatoms and adatom-dimers. 

To begin, we assume we have in hand a magnetic system with an initial charge density $n_0(\vec{r})$. Its ground state magnetization ($m_z(\vec{r})$) pointing along, say, the $z$-direction experiences a modification induced by a small time-dependent 
external transverse magnetic field $B_{ext}(\vec{r};t)$. The result is an 
induced transverse magnetization $m_{x,y}(\vec{r};t)$ localized in the ($xy$) plane perpendicular to the direction $z$. To describe the transverse magnetization, we begin by calculating the frequency dependent Kohn-Sham transverse susceptibility or $\chi_0$ which may be expressed in the form 
\beq
\chi^{ij}_{0}(\vec{r},\vec{r}';\omega)&=&-\frac{1}{\pi}\int dz f(z)
(G^{\downarrow}_{ij}(\vec{r},\vec{r}';z+\omega)
\mathrm{Im}G^{\uparrow}_{ji}(\vec{r}',\vec{r};z)\nonumber\\
&+& \mathrm{Im}G^{\downarrow}_{ij}(\vec{r},\vec{r}';z)
G^{-\uparrow}_{ji}(\vec{r}',\vec{r};z-\omega))\label{chi0}
\eeq where $f(z)$ is the Fermi distribution function, 
$G$ and $G^-$ represent the retarded and advanced 
one particle Green functions connecting atomic sites $i$ and $j$ and $\mathrm{Im}G=-\frac{i}{2}(G-G^-)$. 

A comment on the notation is in order. In general, the point $\vec{r}$ is in 
unit cell $i$, and $\vec{r}'$ is in unit cell $j$. These vectors are measured from the center of their respective unit cells. Thus, if we wish to describe these two points with respect to a master origin $O$, we will describe the 
notation $\vec{r} + \vec{R}_j$ and $\vec{r}' + \vec{R}_j$, respectively 
where $\vec{R}_{i,j}$ are vectors from $O$ to the center points of cell $i$, $j$. With this convention in mind, the single particle Green function, often described as $G(\vec{r}+\vec{R}_i,\vec{r}'+\vec{R}_j,z)$, will here be described as $G_{ij}(\vec{r},\vec{r}';z)$, a notation that is very convenient 
when the KKR scheme we employ is utilized.

To derive our criterion for choosing an effective $U$, our interest is in 
the static form of the Kohn-Sham susceptibility. 
At $\omega=0$, the expression in Eq.~\ref{chi0} reduces to the usual form of the static magnetic susceptibility:
\beq
\chi^{ij}_{0}(\vec{r},\vec{r}';0)&=&\frac{i}{2\pi}\int dz f(z)
(G^{\downarrow}_{ij}(\vec{r},\vec{r}';z)
 G^{\uparrow}_{ji}(\vec{r}',\vec{r};z)\nonumber\\
&-& G^{-\downarrow}_{ij}(\vec{r},\vec{r}';z)
G^{-\uparrow}_{ji}(\vec{r}',\vec{r};z))\label{chi0_2nd}
\eeq

Our first step it to multiply both sides of Eq.~\ref{chi0_2nd} by  $B_{eff}^j(\vec{r}';\omega=0)$ and then we integrate over
$\vec{r}'$ within the atomic site $j$ and sum up over all sites $j$:
\beq
\sum_j\int d\vec{r}' \chi^{ij}_{0}(\vec{r},\vec{r}';0)B^j_{eff}(\vec{r}';0)&=&\frac{i}{2\pi}\int dz f(z) \sum_j \int d\vec{r}'\\\nonumber
&&(G^{\downarrow}_{ij}(\vec{r},\vec{r}';z)B^j_{eff}(\vec{r}';0)
   G^{\uparrow}_{ji}(\vec{r}',\vec{r};z) \nonumber\\
&-& G^{-\downarrow}_{ij}(\vec{r},\vec{r}';z)B^j_{eff}(\vec{r}';0)
G^{-\uparrow}_{ji}(\vec{r}',\vec{r};z))\label{chi0bis}
\eeq
$B_{eff}$ is given by the difference between the potentials of each 
spin channel ( $V_{\downarrow}-V_{\uparrow}$).

We next use an identity derived in the Appendix that relates the Green function for a given spin channel, say 
$\uparrow$, to the Green function of the opposite spin channel through 
the potential difference $B_{eff}$:
\beq
G^{\uparrow}_{ii}(\vec{r},\vec{r};z)
&=&G^{\downarrow}_{ii}(\vec{r},\vec{r};z)
+\sum_j\int d\vec{r}'G^{\downarrow}_{ij}(\vec{r},\vec{r}';z)B^j_{eff}(\vec{r}';0) 
G^{\uparrow}_{ji}(\vec{r}',\vec{r};z)\label{identity}
\eeq
or
\beq
G^{\uparrow}_{ii}(\vec{r},\vec{r};z)
 - G^{\downarrow}_{ii}(\vec{r},\vec{r};z) &=&
\sum_j \int d\vec{r}'G^{\downarrow}_{ij}(\vec{r},\vec{r}';z)B^j_{eff}(\vec{r}';0) 
G^{\uparrow}_{ji}(\vec{r}',\vec{r};z)
\eeq
Similar relations but written differently have been already used for example in Refs.~\cite{LKAG,muniz}. 

Thus Eq.~\ref{chi0bis} becomes:
\beq
\sum_j\int d\vec{r}' \chi^{ij}_{0}(\vec{r},\vec{r}';0)B^j_{eff}(\vec{r}';0)&=&\frac{i}{2\pi}\int dz f(z) \nonumber\\
&&(G^{\uparrow}_{ii}(\vec{r},\vec{r};z)
 - G^{\downarrow}_{ii}(\vec{r},\vec{r};z)\nonumber\\
&-&G^{-\uparrow}_{ii}(\vec{r},\vec{r};z)
 + G^{-\downarrow}_{ii}(\vec{r},\vec{r};z))
\eeq
which is the same as 
\beq
\sum_j\int d\vec{r}' \chi^{ij}_{0}(\vec{r},\vec{r}';0)B^j_{eff}(\vec{r}';0)&=&-\frac{1}{\pi}\int dz f(z) \nonumber\\
&&(\mathrm{Im}G^{\uparrow}_{ii}(\vec{r},\vec{r};z)
 - \mathrm{Im}G^{\downarrow}_{ii}(\vec{r},\vec{r};z))
\eeq
One can recognize that the right-hand side of the previous equation is simply  $m^i_z(\vec{r};0)$. Thus, we obtain the final form of an important sum rule:
\beq
\sum_j\int d\vec{r}' \chi^{ij}_0(\vec{r},\vec{r}';\omega=0)B^j_{eff}(\vec{r}';\omega=0)
&=& m^i_z(\vec{r};\omega=0)
\label{sumrule}
\eeq 
 
We remark that within the empirical tight-binding scheme, a statement equivalent to Eq.\ref{sumrule} is found in Ref.~\cite{muniz}.

The Kohn-Sham susceptibility $\chi^{ij}_0(\vec{r}t,\vec{r}'t')$ can be 
expanded in terms of real spherical harmonics, $Y$ and when this is done it can be expressed as a sum over angular momenta 
$L$, $L_1$, $L_2$ and $L_3$ as $\sum_{LL_1L_2L_3} \chi^{iLL_1;jL_2L_3}_0(rt,r't') Y_L(\hat{r})Y_{L_1}(\hat{r}')Y_{L_2}(\hat{r}')Y_{L_3}(\hat{r})$. This follows 
since $\chi_0$ is a convolution  
of single particle Green functions (see Eq.~\ref{chi0}). Consequently, within the atomic sphere approximation (ASA) and assuming a spherical magnetic field ${m^i_{x,y}(\vec{r}t)}={m^i_{x,y}(rt)}$, $m^i_z(\vec{r}t)=m^i_z(rt)$ and $B^j_{ext}(\vec{r'}t)=B^j_{ext}(r't)$, Eq.~\ref{sumrule} reads:
\beq
\sum_{j}\int d{r}' \sum_{LL_1L_2L_3}Y_L(\hat{r})Y_{L_3}(\hat{r})\chi^{iLL_1;jL_2L_3}_0({r},{r}';0)B^j_{eff}({r}';0) &\times&\nonumber \\ 
\int d\hat{r}'Y_{L_1}(\hat{r}')Y_{L_2}(\hat{r}')
&=& m^i_z({r};0)
\label{sumrule0}
\eeq 
If one integrates both sides of the previous equation over $d\hat{r}$ and uses $\int d\hat{r} Y_L(\hat{r})Y_{L'}(\hat{r}) = \delta_{LL'}$ one finds:
\beq
\sum_{j}\int d{r}' \sum_{LL_1} \chi^{iLL_1;jL_1L}_0({r},{r}';0)B^j_{eff}({r}';0)
&=& 4\pi m^i_z({r};0)
\label{sumrule00}
\eeq 

If we define
\beq U^j({r'})=\frac{B^j_{eff}({r'};0)}{4\pi m^j_z({r'};0)}\label{Usumrule2}\eeq that is the usual form for the effective $U$ that enters Eq.~\ref{dyson} as generated from 
the Adiabatic Local Spin Density Approximation given in the upcoming section, 
then Eq.~\ref{sumrule00} can be rewritten as: 
\beq
\sum_{j}\int d{r}' \sum_{LL_1} \chi^{iLL_1;jL_1L}_0({r},{r}';0)
 m^j_z(r';0) U^j(r')
&=&  m^i_z({r};0)
\eeq 
or as
\beq
\sum_{j}\int d{r}' \Gamma^{ij}(r,r') U^j(r')
&=&  m^i_z({r};0)\label{sumrule0_bis}
\eeq 
with $\Gamma^{ij}(r,r')=\sum_{LL_1}\chi^{iLL_1;jL_1L}_0({r},{r}';0) m^j_z(r';0)$.

In matrix notation, Eq.~\ref{sumrule0_bis} can be expressed as:
\beq
{\Gamma} \vec{{U}}
&=&  \vec{m}_z
\eeq 
which provides a means of calculating of $U$:
\beq
\vec{U}=\Gamma^{-1}\vec{m}_z
\label{Usumrule}
\eeq
Eq.~\ref{Usumrule} allows us to generate $U$ through knowledge of only the 
ground state magnetization and the Kohn-Sham susceptibility $\chi_0$. An 
analysis of Eq.~\ref{dyson} shows that in the absence of an external magnetic field parallel to the $z$-direction the full dynamic susceptibility $\chi$ will have a pole at zero frequency, if in fact $U$ is generated from Eq.~\ref{Usumrule}. Thus, by this scheme we generate an effective $U$ compatible with the Goldstone theorem. Stated otherwise, the correct $U$ is the one 
with the lowest eigenvalue of the denominator of Eq.\ref{dyson} associated with the magnetic moments as components of the eigenvectors. In the following 
we shall show through explicit calculation that the prescription in Eq.~\ref{Usumrule} can be applied to clusters of moment bearing ions which consists of dissimilar atoms.

\section{The Master Dyson Equation within TD-DFT}
Let us briefly derive the master Dyson equation which leads to Eq.~\ref{dyson} within the TD-DFT. By applying a linear variational approach, one assumes similar initial conditions as the ones in the previous section: {\it i.e.} a magnetic system with an initial charge density $n_0(\vec{r})$, a 
magnetization pointing along the $z$-direction and an exciting time-dependent transverse magnetic field $B_{ext}(\vec{r};t)$ with small magnitude that 
allows us to use linear response theory. The result is an 
induced transverse magnetization localized in the ($xy$) plane perpendicular to the direction $z$. The art of TD-DFT is to 
relate and connect the induced transverse magnetization  $m_{x,y}(\vec{r};t)$  to the externally applied magnetic field. The dynamic susceptibility we seek may be expressed as a functional derivative of the transverse moment with respect to the external field, evaluated at zero external field:  
\beq
\chi^{ij}(\vec{r}t,\vec{r}'t')=\frac{\delta m^i_{x,y}[B_{ext}](\vec{r}t)}{
\delta B^j_{ext}(\vec{r}'t')}\bigg|_{B_{ext}=0,n_0}\label{chi1}
\eeq
where $\chi$ is the response function we seek. In regard to the superscripts $i$, $j$ and the definition of the vectors $\vec{r}$, $\vec{r}'$ see 
the remarks after Eq.~\ref{chi0}. The convention we use here is the same as 
that employed for the single particle Green function.

Within the atomic sphere approximation (ASA) and assuming once more an 
applied magnetic field with spherical symmetry within the unit cell we may write
\beq
 m^i_{x,y}(\vec{r}t)&=&\sum_j\int d\vec{r}'\int dt'\chi^{ij}(\vec{r}t,\vec{r}'t') B^j_{ext}(\vec{r}'t'),
\eeq 
Upon resorting to the spherical harmonic expansion discussed above, this becomes
\beq
m^i_{x,y}({r}t)&=&\sum_j\int d\vec{r}'\int dt'\sum_{LL_1;L_2L_3}\chi^{iLL_1;jL_2L_3}({r}t,{r}'t')\times\nonumber\\
&&Y_L(\hat{r})Y_{L_1}(\hat{r}')Y_{L_2}(\hat{r}')Y_{L_3}(\hat{r}) B^j_{ext}({r}'t')
\eeq 
where $r$ and $r'$ are the magnitude of the vectors $\vec{r}$ 
and $\vec{r}'$.

If we integrate both sides of the previous equation over $d\hat{r}$ we find:
\beq
4\pi  m^i_{x,y}({r}t)=\sum_j \int dr'\int dt'\sum_{LL_1}\chi^{iLL_1;jL_1L}({r}t,{r}'t')
 B^j_{ext}({r}'t')
\eeq

Thus the functional derivative given by Eq.~\ref{chi1} could be simplified to 
\beq
\overline{\chi}^{ij}({r}t,{r}'t')=4\pi\frac{\delta m^i_{x,y}[B_{ext}]({r}t)}{
\delta B^j_{ext}({r}'t')}\bigg|_{B_{ext}=0,n_0}
\eeq
where we define $\overline{\chi}^{ij} = \sum_{LL_1}\chi^{iLL_1;jL_1L}$. The same procedure is repeated for the magnetic response function $\chi_0$ of the Kohn-Sham non interacting system which involves not only $B_{ext}$ but 
$B_{eff}$ as well\cite{gross1}; As mentioned previously, $B_{eff}$ is the magnetic part of the 
effective Kohn-Sham potential ($V^{\downarrow}_{eff}-V^{\uparrow}_{eff}$).
After a Fourier transform with respect to time we obtain a form that 
maps our calculation onto the same structure employed many years ago by 
 Lowde and Windsor~\cite{lowde}. This remains often used in recent 
tight-binding simulations of magnetic excitations~\cite{muniz,costa} where it 
is found that the scheme accurately reproduces results found through 
use of a more sophisticated description of the Coulomb integrals. 
Our derivation elucidates how the structure introduced by Lowde and 
Windsor emerges from TD-DFT.

We now have
\beq
\overline{\chi}^{ij}(r,r';\omega)&=&\overline{\chi}^{ij}_{0}(r,r';\omega) \nonumber \\
&+& \sum_{kl} \int dr''\int dr'''
\overline{\chi}^{ik}_{0}(r,r'';\omega) {U^{kl}(r'',r''';\omega)} \overline{\chi}^{lj}(r''',r;\omega)
\eeq 
where the integrations are only over the magnitude of $\vec{r}$ and 
$\vec{r}'$, with the site labeled matrix function shown. The effective 
Coulomb interaction $U^{ij}(r,r';\omega)$ may be expressed as a functional derivative given by 
\beq
U^{ij}(r,r';\omega)&=&
\frac{\delta B^i_{eff}(r;\omega)}{4\pi\delta m^j(r';\omega)}\bigg|_{B_{ext}=0,n_0}
\label{Uij}
\eeq

Within ALDA prescription of the transverse response of the spin system, Eq.~\ref{Uij} simplifies to~\cite{katsnelson} \beq 
U^{ij}(r,r';\omega)&=&\frac{B^i_{eff}(r; 0)}
{4\pi m^i_z(r;0)}\delta_{r,r'}\delta_{i,j}, \label{Udft} \eeq 

The object in Eq.~\ref{Udft} will be noted as  $U_{\mathrm{DFT}}$ is in the litterature often referred to as the exchange and correlation Kernel $K_{\mathrm{xc}}$. 
This is, it should be noted, exactly the form derived in Eq.~\ref{Usumrule2}   extracted from the sum rule Eq.~\ref{sumrule}.

From Eq.~\ref{Udft}, it is obvious that $U$ could be considered as a  
local exchange splitting divided by the magnetization. 

\section{Calculation of the Kohn-Sham susceptibility}

As shown in Eq.~\ref{chi0}, the Kohn-Sham dynamical susceptibility is a convolution of 
two Green functions. The function  
$\chi_0$ can be separated into a sum of two terms: $I_1$ which involves Green functions that 
are analytical in the same half complex plane, so $I_1$ itself is analytic, 
and then one has $I_2$ which is non analytic~\cite{muniz}. For 
positive frequencies:
\beq
I^{ij}_1(\vec{r},\vec{r}';\omega)&=&\frac{i}{2\pi}\int^{E_F} dz f(z)
\bigg(G^{\downarrow}_{ij}(\vec{r},\vec{r}';z+\omega)G^{\uparrow}_{ji}(\vec{r}',\vec{r};z)\nonumber \\
-&&G^{\downarrow *}_{ji}(\vec{r}',\vec{r};z)G^{\uparrow *}_{ij}(\vec{r},\vec{r}';z-\omega)
\bigg)
\eeq

and
\beq
I^{ij}_2(\vec{r},\vec{r}';\omega)&=&\frac{i}{2\pi}\int^{E_F} dz f(z)
\bigg(-G^{\downarrow}_{ij}(\vec{r},\vec{r}';z+\omega)G^{\uparrow *}_{ij}(\vec{r},\vec{r}';z)\nonumber\\
+&&G^{\downarrow}_{ij}(\vec{r},\vec{r}';z)G^{\uparrow *}_{ij}(\vec{r},\vec{r}';z-\omega)
\bigg)
\eeq

 Such a separation is attractive since 
$I_1$ can be calculated through use of a regular energy contour in the complex plane~\cite{wildberger} with    
a modest k- and energy-mesh. In Ref.~\cite{muniz,costa}, the energy contour consists of a line perpendicular to the real-axis starting at the 
Fermi energy and going to infinity. This is unfortunately not possible with the KKR-method since unwanted core states would then be included. Thus, the 
lower limit of the energy integration is chosen well below the valence band minimum. 
$I_2$ can be calculated along a line parallel 
to the real axis. This requires usually a very substantial numerical effort since a large number of k-points as well as a dense energy mesh are needed. However, it can be shown that the integration is limited to a small energy controlled by $\omega$. In our discussion of spin excitations we are interested in frequencies $\omega$ small compared to bandwidths, so the 
integrations involved in $I_2$ can be carried out readily. The 
computational effort is thus enormously reduced. Upon introducing a variable 
change we may write:
\beq
I^{ij}_2(\vec{r},\vec{r}';\omega)&=&-\frac{i}{2\pi}\int_{E_F-\omega}^{E_F} dz 
G^{\downarrow}_{ij}(\vec{r},\vec{r}';z+\omega)G^{\uparrow *}_{ij}(\vec{r},\vec{r}';z)
\eeq

The use of two different contours can lead to a slightly different treatment of rather similar terms in $I_1$ and $I_2$.  
In order to improve numerical stability, in the present analysis the two terms are arranged so they differ a bit from those 
presented in Ref.~\cite{muniz}. We write 
\beq
I^{ij}_1(\vec{r},\vec{r}';\omega)&=&\frac{i}{2\pi}   \int^{E_F-\omega} dz 
\bigg[f(z)G^{\downarrow}_{ij}(\vec{r},\vec{r}';z+\omega)G^{\uparrow}_{ji}(\vec{r}',\vec{r};z)\nonumber\\
&&-f(z+\omega)G^{\downarrow *}_{ji}(\vec{r}',\vec{r};z+\omega)G^{\uparrow *}_{ij}(\vec{r},\vec{r}';z)
\bigg]\\ \nonumber
&&+\frac{i}{2\pi} \int_{E_F-\omega}^{E_F} dz 
f(z)G^{\downarrow}_{ij}(\vec{r},\vec{r}';z+\omega)G^{\uparrow}_{ji}(\vec{r}',\vec{r};z)
\eeq
The second term on the right hand side of the previous equation can be added to $I_2$ which leads to
\beq
\overline{I}^{ij}_2(\vec{r},\vec{r}';\omega)&=&\frac{i}{2\pi}\int_{E_F-\omega}^{E_F} dz 
G^{\downarrow}_{ij}(\vec{r},\vec{r}';z+\omega)
(G^{\uparrow}_{ji}(\vec{r}',\vec{r};z)
-G^{\uparrow *}_{ij}(\vec{r},\vec{r}';z))
\eeq
while
\beq
\overline{I}^{ij}_1(\vec{r},\vec{r}';\omega)&=&\frac{i}{2\pi}\int^{E_F-\omega} dz 
\bigg(f(z)G^{\downarrow}_{ij}(\vec{r},\vec{r}';z+\omega)G^{\uparrow}_{ji}(\vec{r}',\vec{r};z)\nonumber\\
&&-f(z+\omega)G^{\downarrow *}_{ji}(\vec{r}',\vec{r};z+\omega)G^{\uparrow *}_{ij}(\vec{r},\vec{r}';z)
\bigg)
\eeq
or
\beq
\overline{I}^{ij}_1(\vec{r},\vec{r}';\omega)&=&\frac{i}{2\pi}\int^{E_F} dz 
\bigg(f(z-\omega)G^{\downarrow}_{ij}(\vec{r},\vec{r}';z)G^{\uparrow}_{ji}(\vec{r}',\vec{r};z-\omega)\nonumber\\
&&-f(z)G^{\downarrow *}_{ji}(\vec{r}',\vec{r};z)G^{\uparrow *}_{ij}(\vec{r},\vec{r}';z-\omega)
\bigg)
\eeq
This procedure just outlined is found to be stable and requires to calculate one less Green function. 
Up to now we have considered positive frequencies $\omega$.

Negative frequencies lead to slightly different forms of $I_1$ and $I_2$:
\beq 
\overline{I}^{ij}_1(\vec{r},\vec{r}';\omega)&=&\frac{i}{2\pi}\int^{E_F} dz 
f(z)G^{\downarrow}_{ij}(\vec{r},\vec{r}';z-\omega)
G^{\uparrow}_{ji}(\vec{r}',\vec{r};z)\nonumber\\
&-&f(z-\omega)G^{\downarrow *}_{ji}(\vec{r}',\vec{r};z-\omega)G^{\uparrow *}_{ij}(\vec{r},\vec{r}';z)
)
%\eeq
\\\mathrm{and} \nonumber\\
%\beq
\overline{I}^{ij}_2(\vec{r},\vec{r}';\omega)&=&\frac{i}{2\pi}\int_{E_F-\omega}^{E_F} dz 
G^{\downarrow}_{ij}(\vec{r},\vec{r}';z-\omega)(G^{\uparrow}_{ji}(\vec{r}',\vec{r};z)
-G^{\uparrow *}_{ij}(\vec{r},\vec{r}';z))
\eeq
These expressions can be evaluated with modest numerical efforts since the required Green functions are the same than those calculated for the susceptibilities at positive frequencies.
 
\section{An Approximate form for the Single Particle Green functions\label{section_projection}}
The Green functions are provided by the KKR-GF method~\cite{KKR}: 
\beq
G_{ij}(\vec{r},\vec{r}';z)&=&\sum_{LL_1}-i\sqrt{z}R^{iL}(\vec{r}_<;z)H^{iL}(\vec{r}_>;z)\delta_{ij,LL_1} + 
R^{iL}(\vec{r};z)G^{iL,jL_1}_B(z)R^{jL_1}(\vec{r}';z)\label{gf-kkr}
\eeq
where $G_B$ is the structural Green function. Here  
 the  
regular $R$ and irregular $H$ solutions of the Schr\"odinger equation are energy dependent, and this makes the calculation of $\chi_0$ in Eq.~\ref{dyson} tedious and lengthy. Thus, instead of using Eq.~\ref{gf-kkr} while evaluating $\chi_0$, we introduce the following simplification that captures the 
physics central to the systems of interest to us.

  In its spectral representation, the 
Green function is given by
\beq
G_{ij}(\vec{r},\vec{r}';z) &=& \sum_{\vec{k}}\sum_{LL_1}
\frac{\alpha^i_L(E_{\vec{k}}) R^i_L(\vec{r};E_{\vec{k}}) 
      \alpha^{j*}_{L_1}(E_{\vec{k}})
R^{j*}_{L_1}(\vec{r}';E_{\vec{k}})}
     {z-E_{\vec{k}}}
\eeq
where $ R^i_L(\vec{r};E_{\vec{k}})$ is a suitably normalized solution 
of the Schr\"odinger equation within the unit cell $i$.

Various Ansatz can be proposed to simplify the previous form. Instead of 
working with the energy dependent wave functions, one could use an energy linearized form of the wave function as done, for example, in the Linear Muffin Tin Orbital method~\cite{LMTO} or in the Full Potential Linearized Augmented Plane Waves method~\cite{FLAPW}. Our Ansatz expresses the Green functions in terms of energy independent wave functions $\phi$ such that:
\beq
G_{ij}(\vec{r},\vec{r}';z) &\sim& \sum_{\vec{k}}\sum_{LL_1} 
\frac{\beta^i_L(E_{\vec{k}}) \phi^i_L(\vec{r})
      \beta^{j*}_{L_1}(E_{\vec{k}}) \phi^{j*}_{L_1}(\vec{r}')}
     {z-E_{\vec{k}}}
\eeq
or
\beq
G_{ij}(\vec{r},\vec{r}';z) &\sim&\sum_{LL_1} \phi^i_L(\vec{r})
\overline{G}^{LL_1}_{ij}(z) \phi^{j*}_{L1}(\vec{r}')
\eeq
with
\beq
\overline{G}^{LL_1}_{ij}(z) &=& \sum_{\vec{k}} \frac{\beta^i_L(E_{\vec{k}})\beta^{j*}_{L_1}(E_{\vec{k}})}{z-E_{\vec{k}}}
\eeq
Note that after modifying the wave functions we naturally replaced the amplitude $\alpha$ by a different one ($\beta$).

Since our KKR-GF method generates the full Green function as given in Eq.~\ref{gf-kkr}, one could calculate $\overline{G}^{LL_1}_{ij}(z)$ 
from 
\beq
\overline{G}_{ij}^{LL_1}(z)&=&\frac{\int\int d\vec{r}d\vec{r}'\phi^{iL*}(\vec{r})G_{ij}(\vec{r},\vec{r}';z)\phi^{jL_1}(\vec{r}')}
{\int dr\phi^{iL*}(r)\phi^{iL}(r) \int d{r}'\phi^{jL_1}({r}')\phi^{jL_1*}({r}')}\label{Gll}
\eeq
where on the right hand side of Eq.~\ref{Gll} we insert the full 
KKR Green function displayed in Eq.~\ref{gf-kkr}.

The terms in the denominator are normalization 
factors. Thus, instead of working with $\phi^{iL}(\vec{r})$ we introduce  
\beq
\psi^{iL}({r})&=&\frac{\phi^{iL}({r})}
{\bigg(\int d{r}\phi^{iL*}({r})\phi^{iL}({r})\bigg)^{\frac{1}{2}}}
\eeq
 where 
we choose $\phi^{iL}(r)=R^{id}(r;E_F)$, {\it i.e.}, 
the $d$-regular solution of the Schr\"odinger equation. This is appropriate for the calculation of the 
$d$-block of the susceptibility.  We propose here an expansion in terms of energy independent $d$ like wave functions we choose to be the regular solutions of KKR-GF theory evaluated at the Fermi energy. Our focus is on low energy excitations of 3$d$ moments so as we shall see below this choice is appropriate.

Within the KKR-representation of the Green function 
$\overline{G}_{ij}(z)$ is evaluated from:
\beq
\overline{G}^{LL_1}_{ij}(z)&=&\sum_{L_2L_3}\bigg(
-i\sqrt{z}\int_0^{r_{ws}}d\vec{r'}H^{iL_2}(\vec{r}';z)\psi^{iL}(\vec{r}')\int_0^{r'} d\vec{r}\psi^{iL_1*}(\vec{r})R^{iL_2}(\vec{r};z)\delta_{ij,L_2L_3} \\ \nonumber
&&-i\sqrt{z}\int_0^{r_{ws}}d\vec{r'}R^{iL}_2(\vec{r}';z)\psi^{iL}(\vec{r}')\int_{r'}^{r_{ws}} d\vec{r}\psi^{iL_1*}(\vec{r})H^{iL_2}(\vec{r};z)\delta_{ij,L_2L_3}\\ \nonumber
&&+\int_0^{r_{ws}} d\vec{r}\psi^{iL*}(\vec{r})R^{iL_2}(\vec{r};z)G_B^{iL_2,jL_3}(z)\int_0^{r_{ws}} d\vec{r'}R^{jL_3}(\vec{r}';z)\psi^{jL_1}(\vec{r}')\bigg)\label{gfprojected}
\eeq

where $r_{ws}$ stands for Wigner-Seitz radius.

\section{The Final Dyson Equation\label{section_finaldyson}}
Assuming the expansion in terms of energy independent wave functions described previously, the final Dyson equation simplifies after some straightforward 
algebra into a strictly site dependent equation  
\beq
\overline{\overline{\chi}}=\overline{\overline{\chi}}_0 + 
\overline{\overline{\chi}}_0\overline{U}\overline{\overline{\chi}}
\eeq
where the d-block of the dynamical susceptibility is given by
\beq 
\overline{\chi}_0^{ij}(r,r';\omega)&=& \psi^{id}_{\downarrow}(r)\psi^{id*}_{\uparrow}(r)
\overline{\overline{\chi}}_0^{ij}(\omega)\psi^{jd*}_{\downarrow}(r')\psi^{jd}_{\uparrow}(r')\label{suscep_dblock}\eeq

 and 

\beq
\overline{U}^i&=&\int_0^{r_{ws}} dr\psi^{id*}_{\downarrow}(r)\psi^{id}_{\uparrow}(r)U^i(r)\psi^{id}_{\downarrow}(r)\psi^{id*}_{\uparrow}(r)
\label{Upractice}
\eeq
Within ALDA, we use Eq.~\ref{Udft} in Eq.~\ref{Upractice} and obtain
\beq
\overline{U}^i&=&\int_0^{r_{ws}} dr\psi^{id*}_{\downarrow}(r)\psi^{id}_{\uparrow}(r)\frac{B^i_{eff}(r; 0)}
{4\pi m^i_z(r;0)}\psi^{id}_{\downarrow}(r)\psi^{id*}_{\uparrow}(r)
\label{Upractice_dft}
\eeq
If we want to use the sumrule we expand the susceptibility given 
in Eq.~\ref{sumrule00} in terms of d-bloch susceptiblity expressed in 
Eq.~\ref{suscep_dblock} and repeat the same procedure used in section~\ref{section_sumrule} to find 
\beq
\vec{\overline{U}}=\overline{\Gamma}^{-1}\vec{M_z}
\label{Upractice_sumrule}
\eeq
as written in matrix notation and $\overline{\Gamma}^{ij}=\overline{\overline{\chi}}_0^{ij}(0)M_z^j$ with 
$M_z^i$, calculated from the 
projection scheme proposed in section~\ref{section_projection}, is the magnetic moment of atom $i$. 
$\overline{U}$
 can be calculated once for every atom 
either from the previous sum rule, 
Eq.~\ref{Upractice_sumrule}, or from Eq.~\ref{Upractice_dft}.
 It can be understood as a Stoner parameter 
and gives once more a justification for the 
approach used by Lowde and Windsor\cite{lowde}: i.e. the effective intra-atomic Coulomb interaction is expressed 
by only one parameter.

\section{Application of the Formalism to Explicit Examples}

\subsection{Single Adatoms}
We choose as an application of the formalism developed above 
the investigation of 3$d$ adatoms and dimers positioned on the 
fourfold hollow sites of 
Cu(001) surface. In this section, we focus on single adatoms. 
The calculations consist of the self-consistent determination of the 
electronic structure of these nanostructures using the usual KKR-GF scheme~\cite{KKR}. Once 
this is done, we generate the Green functions needed to calculate $\chi_0$, for the elements that 
bear a magnetic moments (Cr, Mn, Fe and Co), following 
Eq.~\ref{chi0}. $U$ is calculated either from Eq.~\ref{Upractice_sumrule} or Eq.~\ref{Upractice_dft}. It is convenient to note that for the case of a 
single adatom $i$, Eq.~\ref{Upractice_sumrule} simplifies to 
$U_{\mathrm{i}}=\frac{1}{\chi^{\mathrm{i,i}}_0}$ at $\omega=0$.

\begin{figure}%[ht!]
\begin{center}
\includegraphics*[angle=0,width=1.\linewidth]{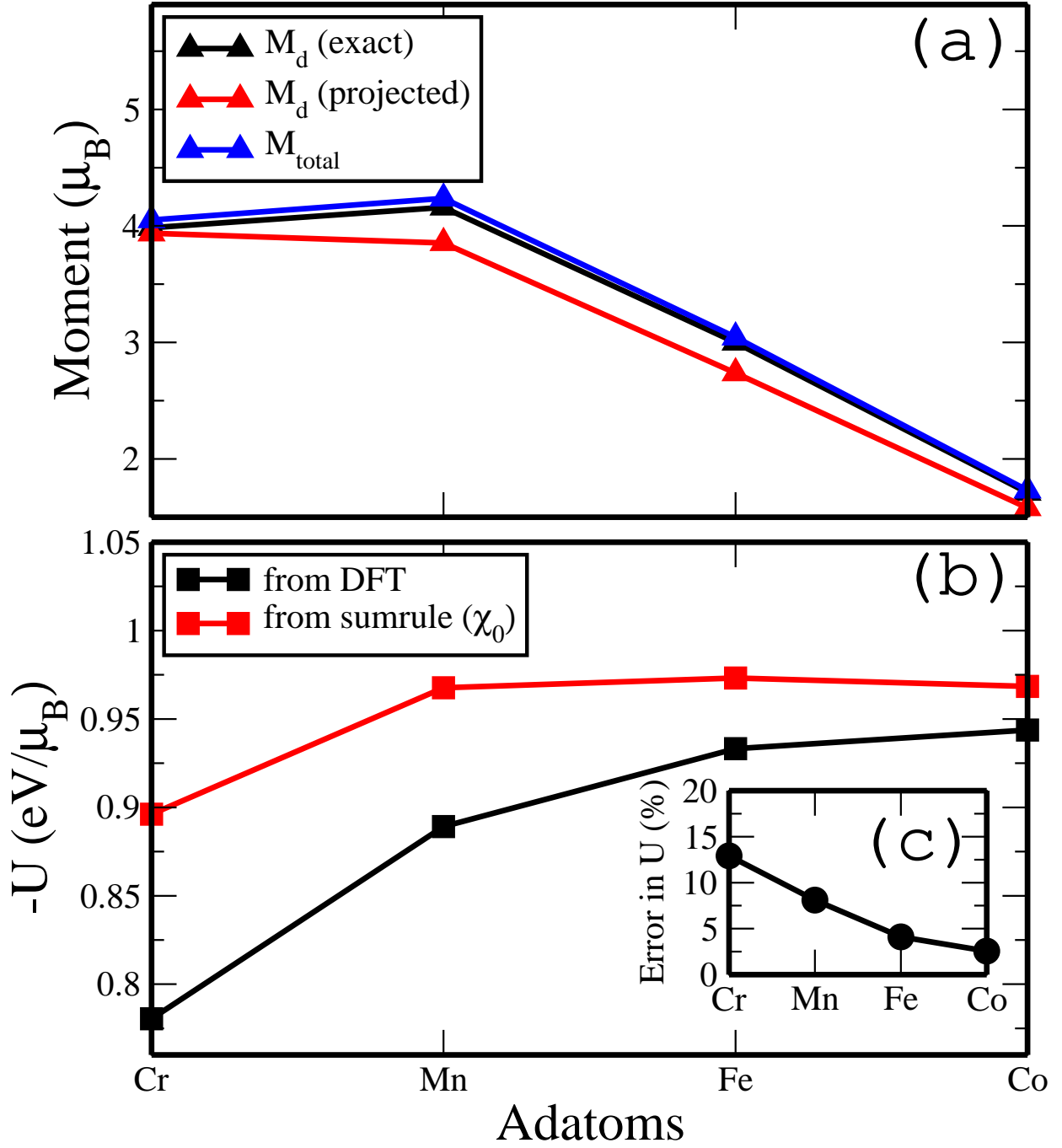}
\end{center}
\caption{(a) Comparison between magnetic moments (in $\mu_B$) of adatoms calculated by the full KKR-GF with those calculated from the proposed projection scheme discussed in the text and in Ref.~\cite{lounis}. Values of $-U$ (eV/$\mu_B$)  are shown in (b) calculated from either from Eq.~\ref{Upractice_sumrule} or from Eq.~\ref{Upractice_dft} while in the insert (c) we plot 
the percentage error defined as the difference between $U_{\mathrm{DFT}}$ and 
$U_{\mathrm{sumrule}}$ divided by $U_{\mathrm{DFT}}$. }
\label{comparison}
\end{figure}

We have already examined the spin dynamics of these systems in Ref.~\cite{lounis} where we 
have shown that the Green functions extracted from our  approach  (Eq.~\ref{gfprojected}) nicely reproduces the magnetic moment of the adatoms as calculated from a full DFT calculation. That this is so is illustrated  in 
Fig.~\ref{comparison}(a). Indeed, interestingly, the $d$-contribution to the total moment 
is, as expected, the most important 
and seems to be nicely reproduced by the projection of the Green functions 
into our choice of wave functions.

 We did not, however, discuss in Ref.~\cite{lounis} the differences between 
values of $U$ calculated from both schemes mentioned previously. In Fig.~\ref{comparison}(b) we show the values of $U$ for the adatoms we 
have investigated. We find values of $U$ very close to $1 eV/\mu_B$ 
for all cases we have studies. Himpsel\cite{himpsel}, in his analysis of a large body of photoemission data on moment bearing 3$d$ ions, has concluded that 
$1 eV/\mu_B$ is a universal value that applies to diverse moment bearing 3$d$ transition metal ions. As discussed in Ref.~\cite{lounis}, $1 eV/\mu_B$ is 
also used commonly ETB calculations~\cite{muniz,costa}. Thus, we are pleased 
to see these values emerge from the scheme set forth here. The relative error or $U$ values generated from density functional theory, as measured by the ratio ($\frac{U_{\mathrm{DFT}}-U_{\mathrm{sumrule}}}{U_{\mathrm{DFT}}}$) 
are depicted in Fig.~\ref{comparison}(c). The error is the highest for Cr-adatom while the lowest is seen for Co. It is interesting that the observed error does not exceed 15$\%$ which is still much lower then what has been estimated by Sasioglu {\it et al.}\cite{sasioglu} while investigating bulk Ni. 

\begin{figure}%[ht!]
\begin{center}
\includegraphics*[angle=0,width=1.\linewidth]{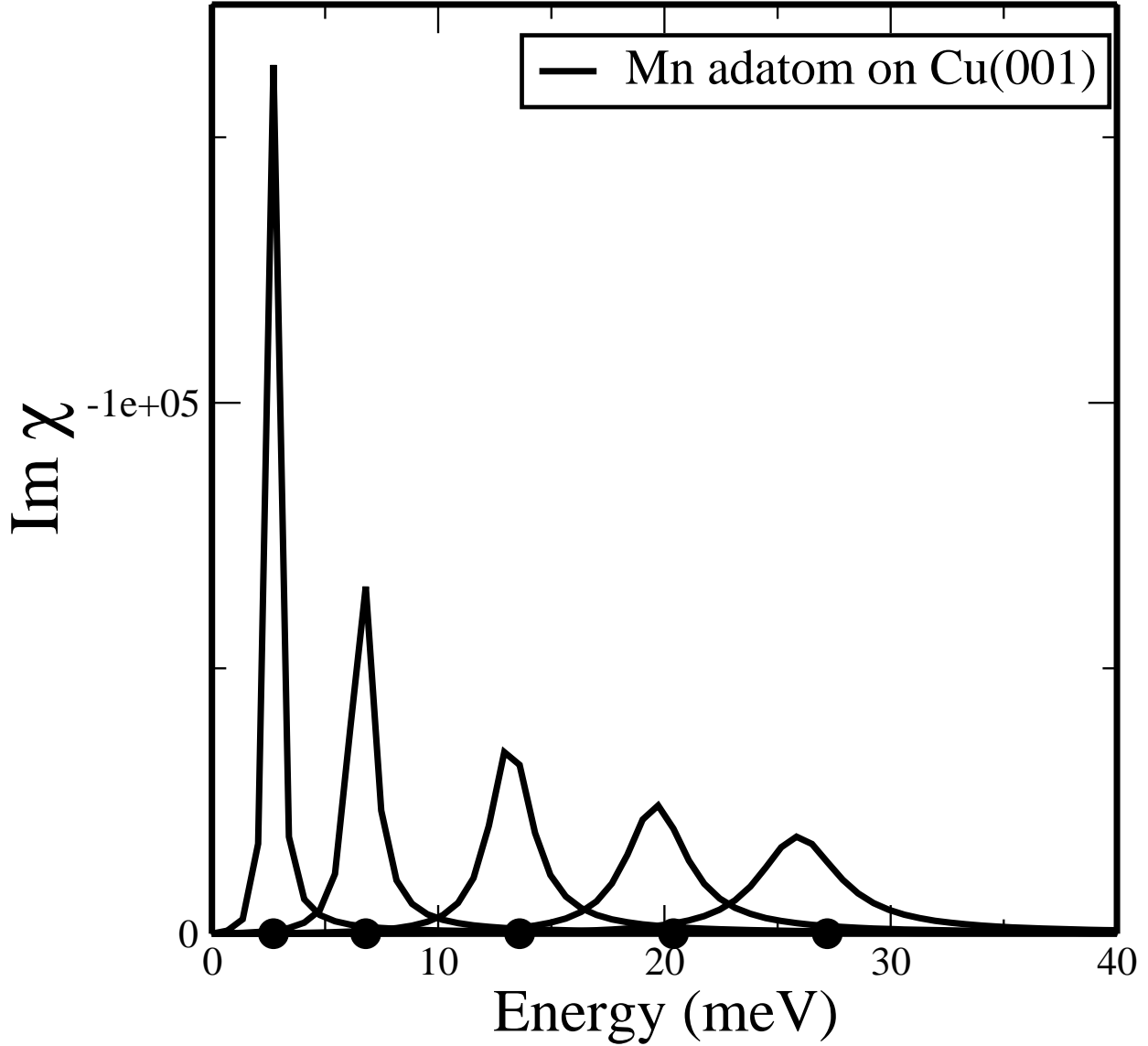}
\end{center}
\caption{Imaginary part of the transverse dynamical magnetic susceptibility 
for a Mn adatom/Cu(001) surface. After applying different DC magnetic fields, resonances 
are obtained and are shifted to higher frequencies by increasing the magnitude of the field. 
The corresponding Zeeman frequency with g=2 for the fields chosen are represented by the black circles. Thus the g shift is negative for this example.}
\label{Mn_susc_bfield}
\end{figure}

In Fig.~\ref{Mn_susc_bfield}, we show examples of the imaginary part of $\chi$ for a Mn adatom 
positioned on the fourfold hollowsite of Cu(001) surface after applying an additional spatially uniform static magnetic field. 
 The imaginary part of $\chi$ describes on the resonant response of the local magnetic moment of Mn-adatom. As required by the Goldstone theorem, a zero frequency resonance is expected when no DC field is applied. We have verified numerically that this feature 
is present, when our method of determining $U$ is employed.
 As soon as a DC 
field pointing along the initial direction of the moment is applied, as discussed many years ago~\cite{lederer}, 
the local response of the moment displays a g shifted Zeeman 
resonance, broadened very substantially by decay of the coherent spin precession to particle hole pairs, whereas the 
total moment of the system precesses with g=2 and zero linewidth. 
Thus, experiments such as STM that are highly localized probes of the dynamic response of the moment see a qualitatively different response than very long wavelength probes such as microwave resonance or Brillouin light scattering. In the latter methods, both g shifts and linewidths have their origin only in terms in the system Hamiltonian that break spin rotation invariance. Examples are spin orbit effects, along with coupling of spins to lattice degrees of freedom.

We see in Fig.~\ref{Mn_susc_bfield} that the resonant frequency scales 
linearly with the applied DC field, as does the width of the structure in 
the local response of the moment. The width of 
the resonances is controlled by the local density of states~\cite{lederer}, and is thus strongly influenced by the 
position of the d levels relative to the Fermi energy.

\subsection{Dimers of Identical Adatoms}
Let us turn to the case of dimers. We consider two identical adatoms each 
adsorbed in nearest neighbor four fold hollow sites on Cu(100). 
 At such distances, their interaction is modest compared to energies 
which characterize the one electron properties of the system.

\begin{figure}%[ht!]
\begin{center}
\includegraphics*[angle=0,width=1.\linewidth]{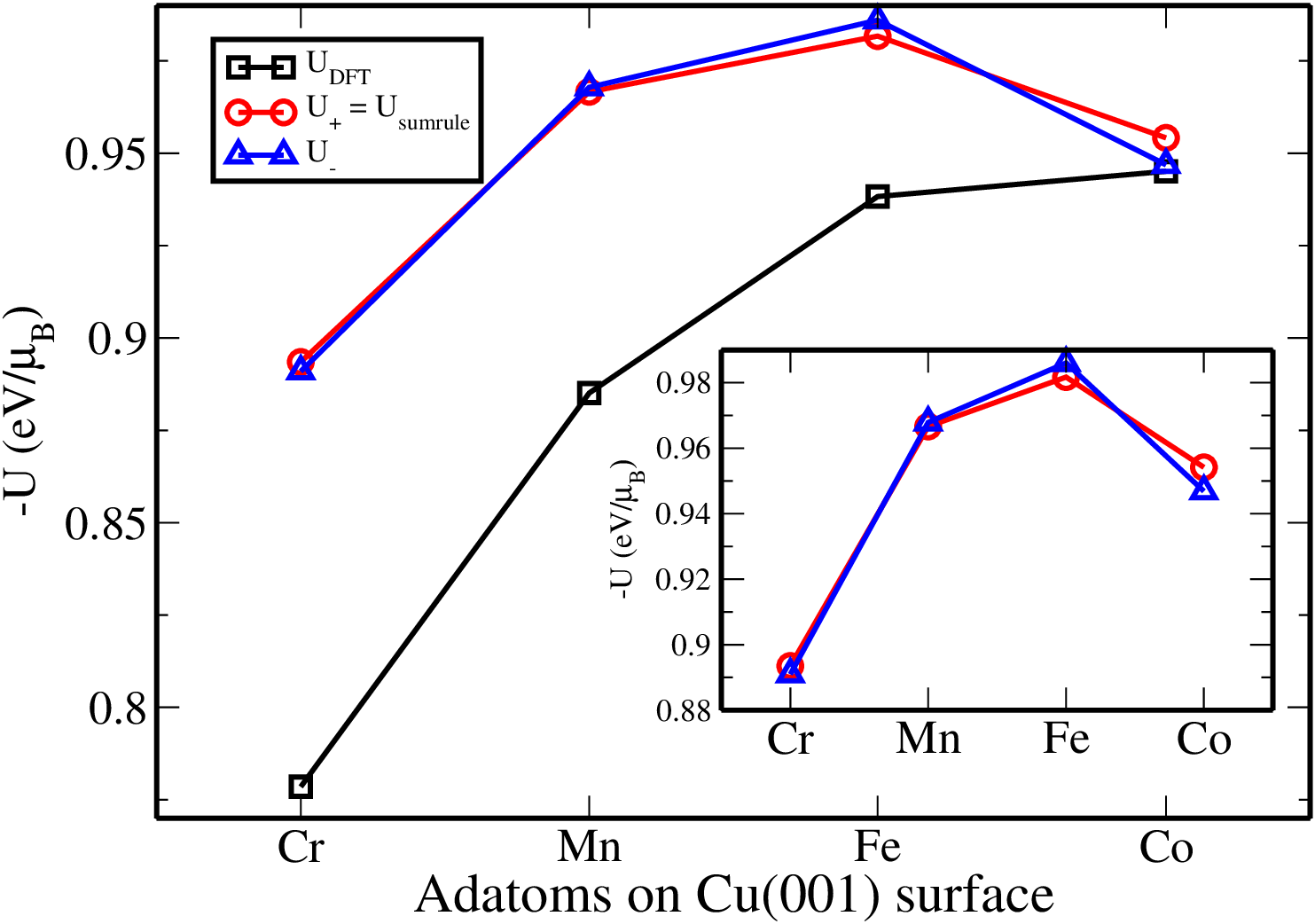}
\end{center}
\caption{Different values of $U$ obtained with different schemes for Cr, Mn, Fe and Co dimers deposited on Cu(001) 
surface. See the discussion in the text for the discussion of the various criteria for choosing $U$.}
\label{U_dimers}
\end{figure}

In Fig.~\ref{U_dimers}, we show effective values of $U$ generated by 
different means of selecting this parameter. The one calculated with 
use of Eq.~\ref{Upractice_dft}, refereed to as $U_{\mathrm{DFT}}$,  is systematically smaller than that which follows from the sum rule in Eq.~\ref{Upractice_sumrule}. 
We saw the same trend in our earlier discussion of single adatoms. 
Of course, if one employs $U_{\mathrm{DFT}}$ in the calculation of the 
dynamic susceptibility the Goldstone theorem is not obeyed. 
We turn next to a discussion the two choices $U_+$ and $U_-$ that appear 
in Fig.~\ref{U_dimers}.

We discuss local dynamic susceptibilities $\chi^{11}$, $\chi^{22}$, 
$\chi^{12}$ and $\chi^{21}$. The superscripts refer to atomic sites 
where the atoms in the dimer are located. The response function $\chi^{ij}$ 
gives the response of the moment at site $i$ in response to a spatially localized field applied to site $j$. So far, everywhere, upper cases were used for $i$ and $j$ site labels in the susceptibility. For the case considered in 
this section, where each atom in the dimer is identical and there is reflection symmetry through the midpoint of the line that connects their 
centers, we have $\chi^{11} = \chi^{22}$ and also $\chi^{21} = \chi^{12}$;  
In the next section we consider a dimer formed from two dissimilar atoms, 
so the equalities just stated do not hold.

The Goldstone theorem requires that in the absence of an externally applied field (and in the absence of spin-orbit coupling) each element $\chi^{ij}$ must have a pole at zero frequency. This is insured if $U$ is such that the determinant $D$ formed from the matrix $1-U\chi_0$ vanishes at zero frequency. 
For our dimer that consists of two identical atoms we have $D=(1-U\chi_0^{11})^2+(U\chi_0^{12})^2$. Upon setting $D=0$, we encounter a difficulty. The criterion yields two acceptable values of $U$, $U_+=(\chi_0^{11}+\chi_0^{12})^{-1}$ and $U_-=(\chi_0^{11}-\chi_0^{12})^{-1}$. 
In Fig.~\ref{U_dimers}, the red curve provides values of $U_+$, for the ions we consider, and the blue curve $U_-$. The two values of $U$ determined by this criterion are quite close to each other, because on the electron volt 
scale the interaction energy between the two moments in the dimer is quite small, as noted above.

One then must address which of the two choices for $U$ discussed in the previous paragraph is the proper physical choice. To see this, we must refine our criterion. For the dimer with two identical atoms, we can make a decision which value of $U$ is the proper choice. If we consider the mode structure of the dimer, there is an acoustical mode wherein the two moments precess in phase, and an out of phase optical mode we shall discuss below. The Goldstone theorem requires the acoustical mode to have zero frequency. Thus, it is the function $\chi^a=\chi^{11}+\chi^{22}+\chi^{12}+\chi^{21}$ that also must have a pole at zero frequency, since this describes the response of the total moment of the dimer to a spatially uniform applied transverse field. For our simple dimer formed from two identical atoms, it is a simple exercise to find an expression for $\chi^a$. One has $\chi^a=(\chi_0^{11}+\chi_0^{12})/[1-U(\chi_0^{11}+\chi_0^{12})]$. Thus for a pole to occur at zero frequency in this response function, we must choose $U=U_+$. 
The sum rule provides us with the same criterion.

For the case of the dimer just considered, it is straightforward to deduce the appropriate choice of $U$ through examination of $\chi^a$. However, for more 
complex arrays of spins the task of choosing $U$ is not simple. Suppose, for 
instance we have $N$ spins in the form of a one dimensional structure or 
possibly an island. From the numerical point of view, one may work with the analog of the determinant $D$ discussed above. 
Exploration of its zeros at zero frequency will yield $N$ possible values of 
$U$. Also if the spin structure consists of dissimilar atoms, each 
atom will be characterized by an appropriate value of $U$. As we shall see in 
the next section, the sum rule allows one to generate appropriate values of the interaction strength for each individual atom in a more complex structure.

We turn next to the description of the spin dynamics of the dimer. 
For the dimer, we expect two resonances, an acoustical mode located obviously at $\omega = 0$ and an optical mode at positive or negative frequencies.  
 In general, the appearance of negative frequency modes in the dynamic susceptibility signal an instability of an assumed ground state. In the studies presented here, we assume a ferromagnetic ground state for the dimer. 
The appearance of a negative frequency optical mode is a signal that the atoms in the dimer are coupled antiferromagntically, so the ferromagnetic ground state is unstable. Thus the dynamic susceptibility can be used as a probe 
of local stability of assumed structures.

It will be useful and interesting to compare our full dynamical calculations of the response of the dimer with the often used localized spin model, where 
effective exchange interactions are calculated within an adiabatic scheme. 
Such adiabatic scheme has already been used for the investigation of different kind of systems (see {\it e.g.}Refs.~\cite{lounis2,mavropoulos,sipr,klautau}). 
Through adiabatic rotations of the moments,~\cite{LKAG}, we extract an effective exchange magnetic interaction, $J$, by fitting the energy change to the Heisenberg form 
\beq
H&=&-J\vec{e}_1\cdot \vec{e}_{2}\label{heisenberg1}
\eeq where        
$\vec{e}_1$ and $\vec{e}_2$ are unit vectors. By this criterion, we find that the ground state is antiferromagnetic for Cr- ($J=-19.8$ meV) 
and Co-dimers ($J=-14.9$ meV) and 
ferromagnetic for Mn ($J=16.3$ meV) and Fe ($J=30.4$ meV). 
Since the dynamical susceptibility was evaluated through use of  
ferromagnetic state for all the dimers, we expect an optical mode 
at positive frequencies for Mn and Fe dimers and at negative frequencies for Cr and Co dimers. 

\begin{figure}%[ht!]
\begin{center}
\includegraphics*[angle=0,width=1.\linewidth]{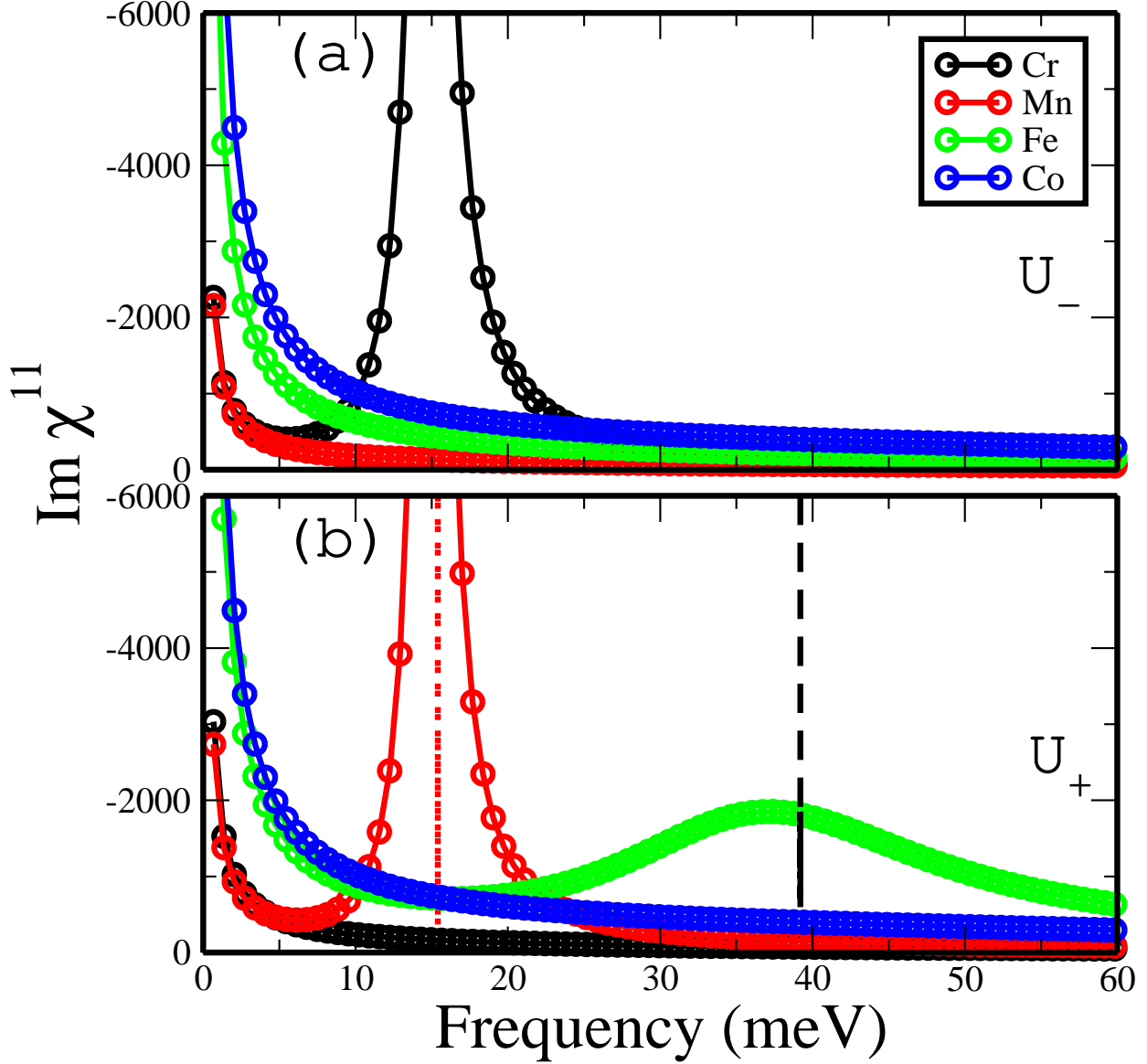}
\end{center}
\caption{local Im$\chi^{11}$ is shown for the four dimers based on: Cr-, Mn-, Fe-, Co- adatoms. To calculate $\chi$ two possible schemes of evaluating are considered: either in (a) using $U_-$ or in (b) using $U_+$. 
It turns out that $U_+$ corresponds to the value obtained from the sumrule 
(Eq.~\ref{Upractice_sumrule}) derived in the text while $U$ that calculated from a simple iterative scheme out of $U_{\mathrm{DFT}}$ would converge to the 
wrong $U$ when investigating Cr and Co dimers. The reason is that, for the latter elements, contrary to $U_+$, $U_-$ is closer to $U_{\mathrm{DFT}}$.
 The optical 
modes, estimated for Mn and Fe from a Heisenberg model, are represented as dashed lines.}
\label{dimers_both_U_nobfield}
\end{figure}

We find that the dynamic susceptibility of the dimer is remarkably 
sensitive to the choice of the effective $U$. We see in Fig.~\ref{U_dimers} 
that numerically the difference between $U_+$ ($=U_{\mathrm{sumrule}}$) and 
$U_-$ is quite small. Yet as illustrated in 
Fig.~\ref{dimers_both_U_nobfield}(a), we show $\mathrm{Im}(\chi^{11})$ calculated with 
the choice $U=U_-$. For all four magnetic ions, the signature of the Goldstone 
mode is evident. For the Cr dimer, we see the clear signature of the optical mode at positive frequency. This suggests that, in contrast to the conclusion 
based on the adiabatic exchange analysis, the ferromagnetic ground state of 
Cr is stable. The optical modes of Mn, Fe all reside at negative 
frequency so for these three the results in Fig.~\ref{dimers_both_U_nobfield}(a) suggest the ferromagnetic ground state is 
unstable. These results are also incompatible with the conclusions based 
on the adiabatic exchange integrals.

In Fig.~\ref{dimers_both_U_nobfield}(b), we show results for $\mathrm{Im}(\chi^{11})$ 
which follow from the choice $U=U_+$. We now have results fully compatible with the conclusion based on the adiabatic exchange analysis. The sum rule has led to the correct selection of the effective $U$.

Within the framework of the Heisenberg model, the optical mode should be an 
eigenmode of the system, and thus it will have zero linewidth. We see in 
Fig.~\ref{dimers_both_U_nobfield}(b) that the optical mode for the 
Fe dimer and the Mn dimer have very substantial width. The origin of this 
broadening is in decay of the optical mode to Stone excitations. 
The itinerant character of the local moments is responsible for this linewidth, which elementary considerations suggest should increase linearly with the frequency of the optical mode. Thus, the linewidth of the optical mode of the Fe dimer is substantially broader than that of the Mn dimer. In 
the ground state, hybridization between 3$d$ states of the adatom and the conduction degrees of freedom on the Cu substrate results in "virtual levels'' 
whose width is in the range of a few hundred meV. At the level of the 
spin dynamics, we see the large broadening of the optical mode as another reflection of the itinerant character of these systems. We note that in 
Spin-Polarized Electron Energy Loss Spectroscopy (SPEELS) studies of spin 
waves in ultrathin films very large linewidths are observed for high frequency, large wave vector modes~\cite{vollner}. The data is in excellent accord with theoretical calculations that assign the large linewidth to the 
damping by decay to Stoner excitations~\cite{costa}, very much as we see 
in the optical modes displayed in Fig.~\ref{dimers_both_U_nobfield}(b).

It is of interest to compare the frequency of the optical modes with 
the prediction of the Heisenberg model, with interspin exchange generated 
adiabatically as discussed above. If one considers two spin exchange 
coupled spins described by the Hamiltonian $-J_s\vec{S}_1\cdot\vec{S}_2$ 
the frequency of the optical mode is easily seen to be $J_s(S_1+S_2)$. 
In Eq.~\ref{heisenberg1}, $\vec{e}_{1,2}$ are unit vectors, so $J_s=J/S_1S_2$. 
Thus, in terms of the effective exchange couplings quoted above, with 
$S_1=S_2=S$ the optical mode frequency is $2J/S$. For the Mn and Fe dimers 
whose optical modes are illustrated in Fig.~\ref{dimers_both_U_nobfield}(b), 
the predicted frequencies are 15.4 meV and 39.2 meV, respectively. The agreement with the optical mode of the Mn dimer is excellent, whereas the full 
dynamical calculation provides a somewhat smaller optical mode frequency for the Fe dimer. As discussed earlier, the coupling between the spin precession of the local moments and the Stoner excitations produces a mode softening 
not incorporated into the localized spin picture\cite{muniz,costa}. This 
coupling is considerably larger for the Fe dimer than the Mn dimer, as 
seen by a comparison of their linewidths.

\subsection{Dimers Formed from Different Adatoms}

We now turn our attention to a lower symmetry spin structure, 
dimers made of different magnetic adatoms. We study the MnFe dimer and 
the FeCo dimer, once again with the magnetic ions sitting in nearest 
neighbor fourfold hollow sites on the Cu(111) surface. Here the two atoms 
do not have the same magnetic moments. Also the effective $U$ is different for 
each atom. In this circumstance it is difficult to envision adjusting 
the values of $U$ by hand to obtain the zero frequency pole in the dynamic 
susceptibility. We have here a circumstance where the sum rule allows us to 
address the problem directly. Notice from Eq.~\ref{Upractice_sumrule} that though its use, we can determine the appropriate value of $U$ for each atom in the dimer. 
Before we discuss imaginary part of the dynamical susceptibility let us discuss values of the magnetic moments and $U$'s.
\begin{table}[ht!]
\begin{center}
\caption{\label{dimer_diff}Comparison between magnetic moments (in $\mu_B$) and values of $U$'s (eV/$\mu_B$) for dimers made of different adatoms: MnFe- and FeCo dimers.}
\begin{ruledtabular}
\begin{tabular}{lcc}
                                   & Mn/Fe       & Fe/Co        \\
\hline
 $M_d$: projection model    &  3.85/2.74      &  2.78/1.64  \\
 $M_{total}$                        &4.23/3.06      &  3.13/1.82        \\
 -$U_{\mathrm{DFT}}$ &0.89/0.94      &  0.94/0.95    \\
-$U_{\mathrm{sumrule}}$&0.97/0.98         &  0.98/0.98            
\end{tabular}
\end{ruledtabular}
\end{center}
\end{table}

In Table~\ref{dimer_diff}, the magnetic moments calculated with our projection scheme are shown and compared to the values that follow from the 
full KKR treatment of the ground state. In the first line of Table~\ref{dimer_diff} the moment which appears is the contribution with 
$d$-like symmetry, since this is the portion built into our Ansatz for the 
Green function used to compute the Khon-Sham susceptibility. It is interesting to note the substantial difference between the magnetic moments of two adatoms in the dimer. It is the case here as for the single adatom, the $U$ calculated from Eq.~\ref{Upractice_dft} understimates the value of $U$ needed to realize the Goldstone mode. From Eq.~\ref{Upractice_sumrule}, we may deduce the value $U$, for each of the adatoms in the dimer. We find  
\beq
U_1=\frac{\frac{m_z^2}{m_z^1}\chi_0^{12}-\chi_0^{22}}
{\chi_0^{12}\chi_0^{21}-\chi_0^{11}\chi_0^{22}}
\eeq
and
\beq
U_2=\frac{\frac{m_z^1}{m_z^2}\chi_0^{21}-\chi_0^{11}}
{\chi_0^{12}\chi_0^{21}-\chi_0^{11}\chi_0^{22}}
\eeq

It is interesting that the sum rule gives similar values of $U$ for both atoms in the dimer, and also that $U$ is very close to $1 eV/\mu_B$. That this is 
so is very compatible with the conclusion of Ref.~\cite{himpsel}, which is 
based on an empirical study of photoemission data on 3$d$ transition metal 
ions in diverse environments.

\begin{figure}%[ht!]
\begin{center}
\includegraphics*[angle=0,width=1.\linewidth]{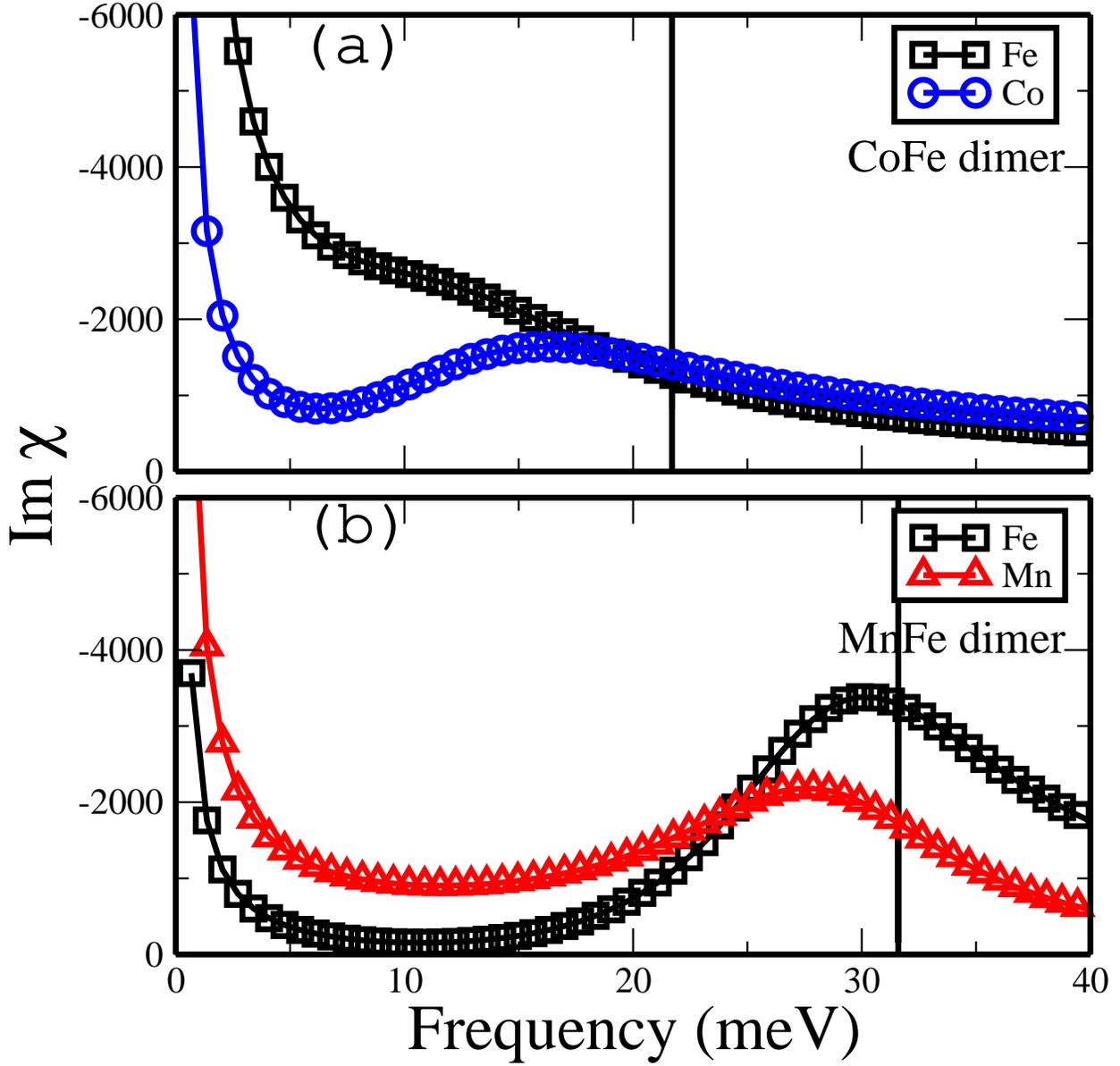}
\end{center}
\caption{local Im$\chi$ for dimers with mixed adatoms are shown in (a) for FeCo dimer and 
in (b) for MnFe dimer. 
Eq.~\ref{Upractice_sumrule} based on the sum rule derived in the text was used to define 
$U$. It is interesting to note the presence of resonances at positive frequencies 
expressing a ferromagnetic ground state for both dimers. Within each dimer, the pics related to every adatom are not located at the same position since the g-shift depends on the nature of the adatom.}
\label{dimer_different}
\end{figure}

The mapping to the previously defined Heisenberg model predicts a ferromagnetic ground state for both dimers investigated. Indeed the magnetic exchange interaction is positive in both cases with 
$J_{\mathrm{MnFe}}=28.1$ meV (Heisenberg frequency 31.6 meV) and $J_{\mathrm{FeCo}}=12.5$ meV (Heisenberg frequency 21.7 meV). This indicates, 
as discussed above, that the imaginary part of the dynamical magnetic susceptibility for every adatom should show a resonance at positive frequencies that is the signature of the optical mode. In  Fig.~\ref{dimer_different}(a) and (b) we plot $\chi^{11}$ and $\chi^{22}$ 
respectively for the FeCo- and MnFe-dimer.

A most striking feature of the results displayed in Fig.~\ref{dimer_different} is that the peak positions in $\chi^{11}$ and $\chi^{22}$ occur at distinctly 
different frequencies. This is particularly clear in Fig.~\ref{dimer_different}(b), where the influence of damping is somewhat more 
modest than in Fig.~\ref{dimer_different}(a). We see that the peak in 
$\chi^{\mathrm{FeFe}}$occurs at 30 meV, whereas that in $\chi^{\mathrm{MnMn}}$ 
is distinctly downshifted to 27 meV.

This behavior is at variance with the Heisenberg description of the 
excitation spectrum of two well defined localized spins. As we have seen, if 
we have two well defined, localized spins coupled together by the 
exchange interaction $-J\vec{e}_1\cdot\vec{e}_2$, the pair has two 
excited states associated with small amplitude motions, the acoustical mode at 
zero frequency (which we see in Fig.~\ref{dimer_different}) and the optical mode at the frequency $2J/S$. Thus, the optical mode peak in the excitation spectrum for each member of the dimer should be at exactly the same 
frequency, in this picture. While the oscillator strength of each peak will 
differ, there is a unique excited state energy of the pair.

\begin{figure}%[ht!]
\begin{center}
\includegraphics*[angle=0,width=1.\linewidth]{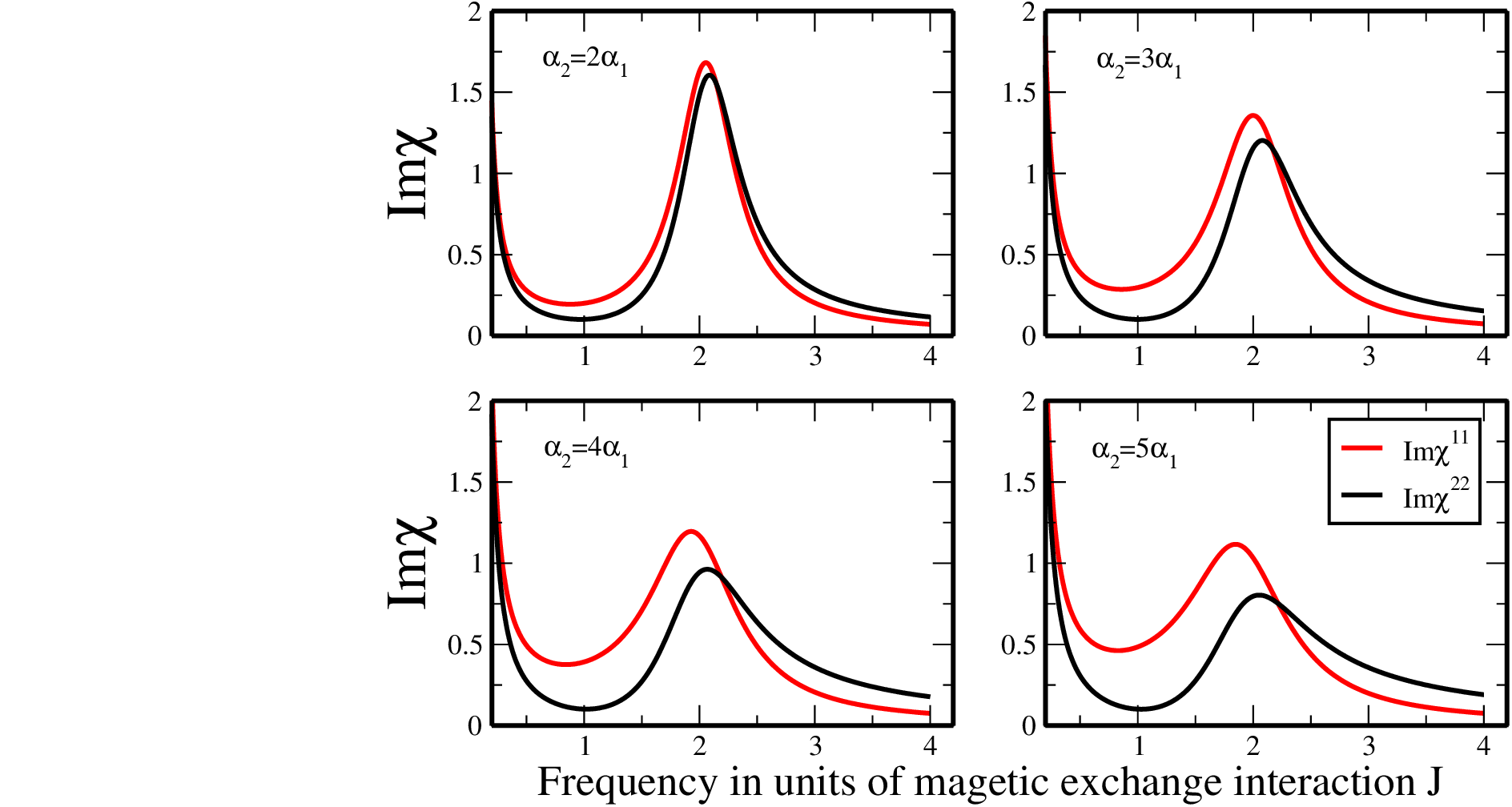}
\end{center}
\caption{The response function Im$(\chi^{11})$ and Im$(\chi^{22})$ 
for two spins of unit length coupled by an exchange interaction of 
strength $J=1$. Here, we mimic Fe and Mn by considering each spin 
coupled to a reservoir that provides a 
damping parameter $\alpha_{1,2}$ (1 for Mn and 2 for Fe) 
whose values are given in the inset.}
\label{dimer_model}
\end{figure}

The shift in the peak positions evident in Fig.~\ref{dimer_different} is  
a consequence of the itinerant nature of the magnetic moments. As each 
moment precesses, as we have seen, the motion is damped heavily by the 
coupling of the moment to the Stoner excitations of the paramagnetic host. 
In the case of the FeMn dimer, the motions of the Fe spin are damped far more 
heavily that those of the Mn spin, as we may appreciated from Fig.1(b) of 
Ref.\cite{lounis}. This has the consequence that the peak in Im$\chi^{\mathrm{MnMn}}$ is dragged down to a frequency somewhat lower than that 
in Im$\chi^{\mathrm{FeFe}}$. We may see this by constructing a toy model 
that consists of two Heisenberg coupled spins, each of which is coupled to 
a reservoir that produces damping $\alpha$ of the form encountered in the 
Landau-Lifschitz-Gilbert equation. The linearized equations of motion for this system reproduces the offset in the peaks evident in Fig.~\ref{dimer_different}(b). We illustrate this in Fig.~\ref{dimer_model} where Im$\chi^{\mathrm{11}}$ and Im$\chi^{\mathrm{22}}$ mimic the imaginary parts of $\chi^{\mathrm{MnMn}}$ and $\chi^{\mathrm{FeFe}}$.  
By increasing the strength of the damping parameter $\alpha_2$ compared to 
$\alpha_1$, we observe a shift to lower energies of the optical mode in 
Im$\chi^{\mathrm{22}}$ ({\it i.e.} Im$\chi^{\mathrm{MnMn}}$). it is striking to 
observe the completely different shape of the optical mode of Mn-spin just by modifying a neighbor. Indeed, by comparing 
the optical mode observed in Im$\chi^{\mathrm{MnMn}}$ we observe also that 
it is much more heavily damped in the mixed dimer MnFe (Fig.~\ref{dimer_different}(b)) than in the pure MnMn dimer (Fig.~\ref{dimers_both_U_nobfield}(b)). The physical reason behind this intriguing behavior is that in the MnFe configuration, the Mn-spin during its  precession feels the magnetic force of the heavily damped Fe-spin which provides more damping on Mn. It would be of great interest to employ STM based spectroscopy to explore the 
response of the two spins in a dissimilar dimer 
such as that just discussed.

\section{Conclusion} We have developed and presented a theory based on TD-DFT and the KKR-GF 
method to extract dynamics magnetic susceptibilities of moment bearing adatoms and adatom dimers on surfaces. In our method, the electronic structure is 
described within an ab-initio scheme with KKR Green functions as the basis. 
Thus, no parameters need to be introduced, as in studies that employ the empirical tight-binding method. As important feature of our approach is that 
it may be implemented with a modest expenditure of computational effort. 
It is thus suitable for exploration of complex magnetic structures on surfaces that contain several magnetic ions. In this paper, we illustrate the method with application to magnetic dimers formed from either identical or dissimilar adatoms.

As discussed above, a difficulty with past TD-DFT studies of spin excitations 
not only on surfaces, but in bulk materials as well is that the effective value of the Hubbard $U$ which emerges from the standard approaches is not compatible with the Goldstone theorem that guarantees that the low lying acoustical spin-excitation has zero frequency. This difficulty has led others to make ad-hoc adjustments in the value of $U$. A feature of the present analysis is the introduction of a sum rule from which proper values of this 
parameter emerge. This eliminates the need for ad-hoc adjustments. It should be remarked that in simple systems, where the analysis can be phrased in terms of 
a single value of the effective $U$, it is not difficult to insure satisfaction of the Goldstone theorem through an ad-hoc correction, though in our view this is an unsatisfactory procedure that compromise the theory at the fundamental level. Additionally, for a multicomponent system, the ad-hoc correction procedure becomes problematic in practice. As 
we see from our discussion of the dimer constructed from two different magnetic ions, our sum rule approach is readily and easily implemented for multi-component systems.

\section*{Acknowledgments}
Research supported by the U. S. Department of Energy through grant No.
DE-FG03-84ER-45083. R.B.M. acknowledges support from CNPq and
FAPERJ, Brazil. S. L. thanks the Alexander von Humboldt Foundation 
for a Feodor Lynen Fellowship and also wishes to thank 
Stefan Bl\"ugel for constant support of this work. The
computations were performed  at the supercomputer JUROPA
at the Forschungszentrum J\"ulich.

\section*{Appendix}
In this appendix we provide a derivation of the useful identity presented in 
Eq.~\ref{identity}.

The Green function $G(z)$ of a Hamiltonian operator $H$ is defined by the 
operator equation
\beq
G=\frac{1}{z-H}
\eeq

If no spin-orbit coupling and non-collinear magnetism are considered, 
the previous equation holds for every spin-channel ($\uparrow$ or $\downarrow$). Thus
\beq
G^{\uparrow(\downarrow)}&=&\frac{1}{z-H^{\uparrow(\downarrow)}}
\eeq

In addition we have:
\beq
z-H^{\downarrow}&=&z-H^{\uparrow}+H^{\downarrow}-H^{\uparrow}
\eeq
that can be multiplyed from both sides from the left by 
$(z-H^{\downarrow})^{-1}$ and from the right by $(z-H^{\uparrow})^{-1}$. 
This leads to
\beq
\frac{1}{z-H^{\uparrow}}&=&\frac{1}{z-H^{\downarrow}}+
\frac{1}{z-H^{\downarrow}}(H^{\downarrow}-H^{\uparrow})\frac{1}{z-H^{\uparrow}}
\eeq  
{\it i.e.}
\beq
G^{\uparrow}&=&G^{\downarrow}+
G^{\downarrow}B_{eff}G^{\uparrow}
\eeq 
where we define  $B_{eff}=H^{\downarrow}-H^{\uparrow}$.


\begin{thebibliography}{99}
\bibitem{future} This is the subject of a future publication including the full 
theory for simulations of inelastic scanning tunneling microscopy experiments. Refs.~\cite{fransson} is useful 
for this matter as well.
\end{thebibliography}
\end{document}